\documentclass[10pt,a4paper,twoside]{article}

	\usepackage{amssymb}
	\usepackage{amsmath}
	\usepackage{bm}
	\usepackage{graphicx}
	\usepackage{subfigure}
	\usepackage{hyperref}
	\usepackage{color}
	\usepackage[font=small,labelfont=bf]{caption}

	\usepackage{booktabs}
	\usepackage{array}
	\usepackage{enumitem}
	\usepackage{cite}
	\usepackage[affil-it,auth-lg]{authblk}
	\usepackage{hyperref}
	\usepackage{eso-pic}% http://ctan.org/pkg/eso-pic

\begin{document}

\title{The Gravothermal Instability at All Scales: From Turnaround Radius to Supernovae \footnote{This paper is based on the talk at the 7th International Conference on New Frontiers in Physics (ICNFP 2018), Crete, Greece, 4--12 July 2018.}}

\author[1]{Zacharias Roupas}
\affil[1]{Department of Mathematics, University of the Aegean, Karlovassi 83200, Samos, Greece} 
	
\date{\vspace{-5ex}}
%\pacs{}
\maketitle

\begin{abstract}
The gravitational instability, responsible for the formation of the structure of the Universe, occurs below energy thresholds and above spatial scales of a self-gravitating expanding region, when thermal energy can no longer counterbalance self-gravity. I argue that at sufficiently-large scales, dark energy may restore thermal stability. This stability re-entrance of an isothermal sphere defines a turnaround radius, which dictates the maximum allowed size of any structure generated by gravitational instability. On the opposite limit of high energies and small scales, I will show that an ideal, quantum or classical, self-gravitating gas is subject to a high-energy relativistic gravothermal instability. It occurs at sufficiently-high energy and small radii, when thermal energy cannot support its own gravitational attraction. Applications of the phenomenon include neutron stars and core-collapse supernovae. I also extend the original Oppenheimer--Volkov calculation of the maximum mass limit of ideal neutron cores to the non-zero temperature regime, relevant to the whole cooling stage from a hot proto-neutron star down to the final cold state. 
\end{abstract}

\section{Introduction}

The self-gravitating, ideal gas is subject to an instability in the Newtonian gravity regime identified in the year $1962$ by Antonov \cite{antonov} {in a thermodynamic framework}. 
It was further elaborated and given an interpretation a few years later by Lynden-Bell and Wood \cite{lbw}, {who coined the name ``gravothermal catastrophe''}. Under perturbations that preserve the energy, a self-gravitating, ideal gas sphere collapses if the total energy value is below a minimum negative threshold for fixed radius or equivalently if the radius value is above a maximum threshold for fixed negative energy. In particular for this Newtonian regime and for intermediate scales ranging from non-compact stars like red-giants to star clusters, stable equilibria do exist only if {$R<0.335GM^2/(-E)$, for $E = E_\text{kinetic}+U_\text{gravity}<0$}, % is the italic necessary for ``GM''? //ZR: Yes, the math mode is necessary for all varibales
while for $E>0$ and assuming the system is bound by external pressure, there exist equilibria for any $R$. {Sormani and Bertin showed recently \cite{2013A&A...552A..37S} that the gravothermal catastrophe is a manifestation of the Jeans instability \cite{Jeans_1902RSPTA.199....1J} under certain boundary conditions, so that they both evidence a single phenomenon, the gravitational instability, in different frameworks, thermodynamic and dynamic.}

In this work, I will complete the analysis of the self-gravitating ideal gas, extending it to all scales, based on some of my contributions to the subject. In the very small scale, equivalently at high energies, general relativity becomes relevant, and especially, the weight of heat of Tolman \cite{Tolman:1930,Roupas_RGI_2018} intervenes, causing a high-energy gravothermal instability that I identified in \cite{Roupas_RGI_2018,Roupas_CQG_RGI_2015}. At big scales, dark energy becomes relevant. It restores thermodynamic stability due to its repulsive nature \cite{Axenides_etal_2012PRD,Axenides_2013NuPhB.871...21A,Roupas_etal_2014PRD}.

\section{The Turnaround Radius}

Consider a slowly-expanding ideal gas sphere passing consecutively through thermal equilibrium states. Assume the gas' constituents interact gravitationally and that the total energy: 
\begin{equation}
	E = E_\text{kinetic}+U_\text{gravity}
		= \frac{3}{2}NkT - \frac{G}{2}\iint \frac{\rho(\vec{r})\rho(\vec{r}\,')}{|\vec{r}-\vec{r}\,'|}d^3\vec{r}d^3\vec{r}\,',
\end{equation} 
remains constant and negative. This is because of the negative total gravitational energy contribution, which we assume to be dominating over the kinetic energy contribution, and therefore, the system is bounded primarily by gravity and not external pressure. The gas cools down during expansion, resulting in the condensation of the central parts because of the diminishing of the thermal energy's ability to counter-balance gravity. Above some threshold maximum radius, the thermal energy can no longer halt the collapse of the core. The system passes from stable equilibria with negative specific heat to unstable ones with positive specific heat, while the core becomes itself self-gravitating, attaining negative specific heat \cite{lbw}. A runaway effect sets in, while heat transfer from the core to the halo cannot be reversed.

\begin{figure}[h]
%\begin{center}
\centering
 	\subfigure[The turnaround radius for $\rho_\Lambda$]{\label{fig:R_turnaround}
 		\includegraphics[scale = 0.0188]{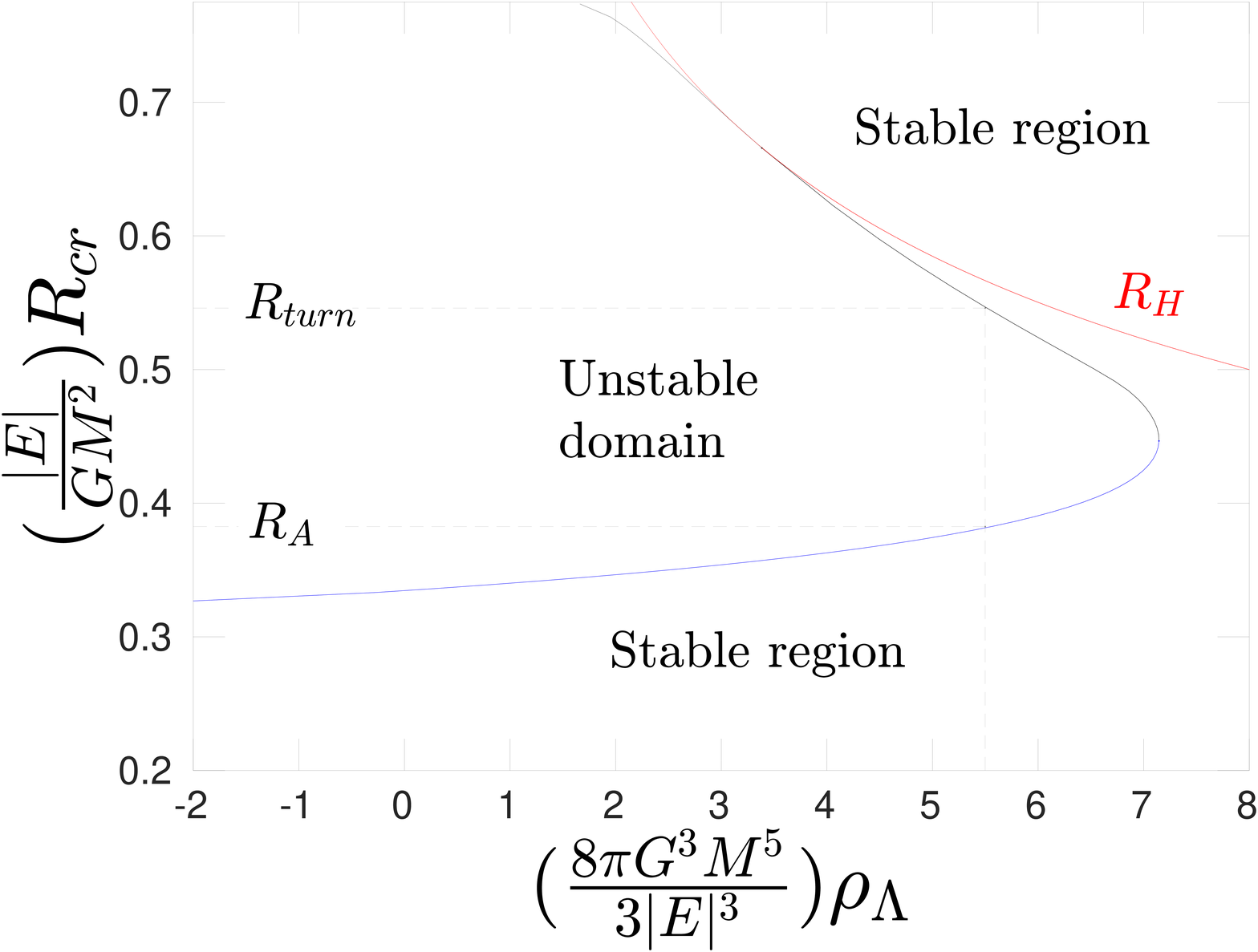} }
 	\subfigure[Quintessence vs. phantom]{\label{fig:Rcr_dark_Q+P}
 		\includegraphics[scale = 0.12]{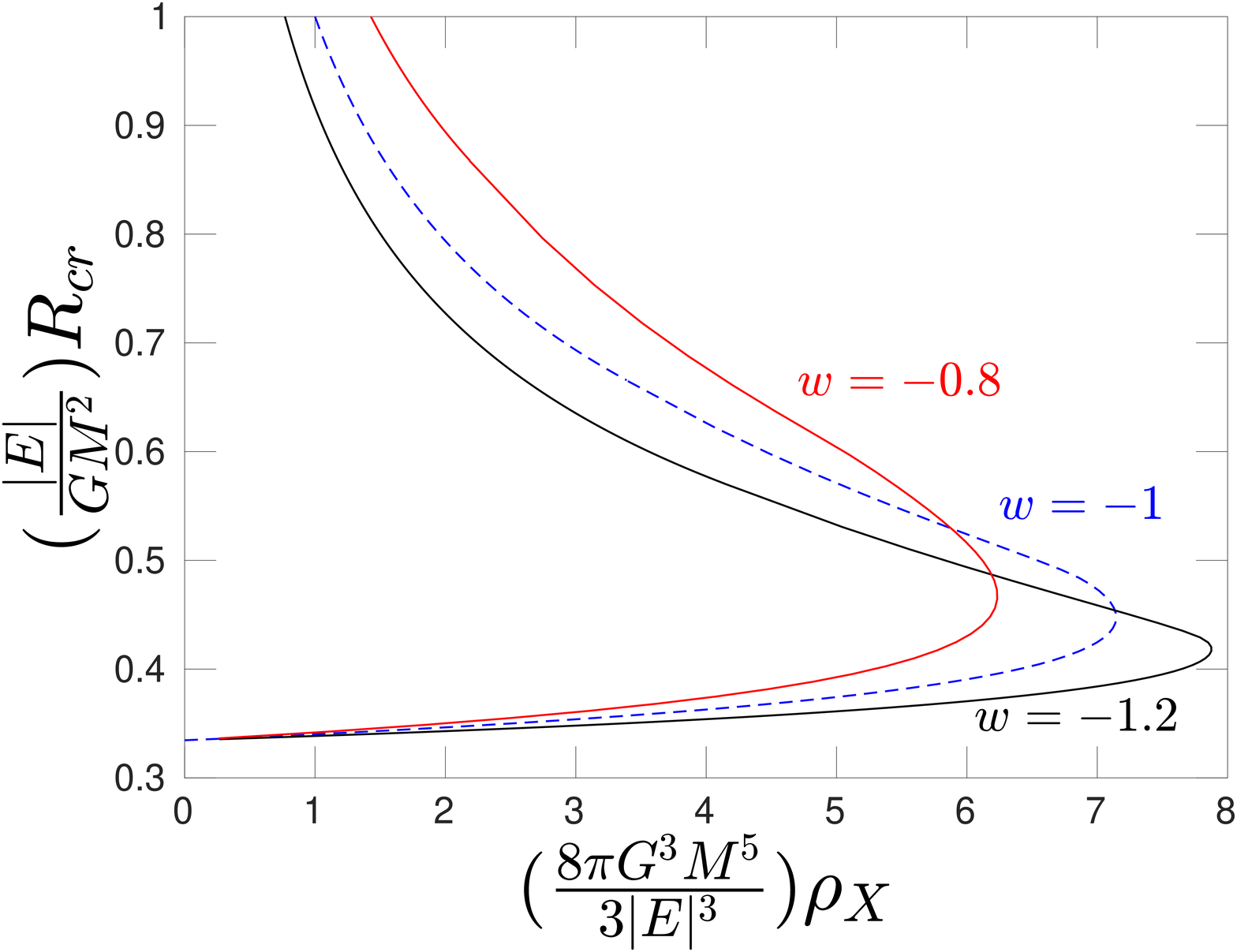} }
	\caption{{(a):} %Figures 1--3: is the italic necessary? should ``Left'' be changed to (a), ``rigth'' be changed to (b)?
	Thermodynamic equilibria of an ideal gas cannot exist below the turnaround radius $R_\text{turn}$ for a fixed energy $E$. We assume here a dark energy density with the equation of state corresponding to the cosmological constant $\Lambda$. Its value can be estimated by this framework if one considers the typical energy of a perturbation and the smaller bound structures observed. Conversely, assuming the $\Lambda$ value is measured by other means, this framework predicts the bigger gravitationally bound structures than may exist in the Universe. At even smaller scales, $R<R_A$ stability is restored, and stable equilibria do exist. Note that the turnaround radius predicted here thermodynamically is stricter and more accurate than the $R_H$ predicted considering only homogeneous equilibria. The figure is taken from \cite{Axenides_etal_2012PRD}.
	{(b):} We consider a quintessential type of dark energy $w=-0.8$ and a phantom type of dark energy $w=-1.2$. It is evident that the unstable domain is reduced for the phantom case. Therefore, quintessence favors the formation of structure w.r.t. the phantom dark energy and the cosmological constant. The figure is taken from \cite{Roupas_etal_2014PRD}.
	\label{fig:dark_energy}}
%\end{center} 
\end{figure}

However, at even bigger scales, dark energy restores the stability of the equilibrium states \cite{Axenides_etal_2012PRD,Axenides_2013NuPhB.871...21A,Roupas_etal_2014PRD}, as depicted in Figure \ref{fig:dark_energy}a. Therefore, a spherical non-homogeneous perturbation will decouple from the expansion and collapse only below a threshold radius, the {turnaround radius} (an incomplete list on the subject includes \cite{Stuchlik_1999PhRvD..60d4006S,Busha_2003ApJ...596..713B,Mizony_2005A&A...434...45M,Pavlidou_2014JCAP...09..020P,Faraoni_2015JCAP...10..013F,Capozziello_2018arXiv180501233C}). This radius estimated by equilibria of an isothermal sphere as we did in~\cite{Axenides_etal_2012PRD,Roupas_etal_2014PRD} is smaller and thus stricter---rendering our estimation more accurate---than the one estimated considering only homogeneous equilibria (e.g., published later \cite{Pavlidou_2014JCAP...09..020P}). 

Assuming an equation of state for dark energy of the form:
\begin{equation}
	P_X = w\rho_X c^2, \; w<0
\end{equation}
we find in addition that quintessential dark energy ($w> -1$) favors the formation of structures with respect to the phantom-type dark energy ($w < -1$), as depicted in Figure \ref{fig:dark_energy}b. 

The diagrams of Figure \ref{fig:dark_energy} are generated by solving numerically the following equation (for a general derivation, see \cite{Roupas_etal_2014PRD}, and for the case $w=-1$, see \cite{Axenides_2013NuPhB.871...21A}):
\begin{equation}
\frac{1}{r^2}\frac{d}{dr}\left( r^2\frac{d}{dr}\phi(r)\right) = 4\pi G\left(\rho_0+(1+w)\rho_X\right) e^{-m\beta(\phi(r)-\phi(0))}
+8\pi G w \rho_X
\end{equation}
for several $R$ boundary values, keeping the total number of particles fixed.
This equation is the modification of the Emden equation due to dark energy, i.e., the Poisson equation for a density corresponding to an isothermal distribution plus the constant energy density and gravitating pressure of dark energy~\cite{Roupas_etal_2014PRD}.

\section{Relativistic Gravothermal Instability}

If the self-gravitating classical ideal gas sphere we considered previously is now compressed at very small scales, general relativity will come into play. The equation that describes now the thermodynamic equilibria is the Tolman--Oppenheimer--Volkoff (TOV) equation:
\begin{equation}
	\label{eq:TOV}
\frac{dP}{dr} = -(\rho + \frac{P}{c^2})\left( \frac{G\hat{M}(r)}{r^2} + 4\pi G \frac{P}{c^2}r\right)
	\left( 1 - \frac{2G\hat{M}(r)}{rc^2} \right)^{-1}.
\end{equation}

We will assume classical particles and therefore the equation of state of a classical ideal gas (see, e.g.,~\cite{Roupas_RGI_2018} for details).

I have shown \cite{Roupas_2013CQG,Roupas_2015CQG_32k9501R,Roupas_RGI_2018} that the TOV equation can be derived by the maximum entropy principle:
\begin{equation}\label{eq:deltaS}
	\delta S - \tilde{\beta} c^2 \delta M + \alpha \delta N = 0. 
\end{equation}
together with the red-shift factor{(!%is the exclmation necessary? should there be a space before (? please check the content here./// ZR: The exclamation serves to notify an unexpected, but important result. You may add a space before the parenthesis but I think this will obscure the meaning of the exclamation mark.
)} assuming only the Hamiltonian constraint, equivalently the expression of the proper spatial volume $d^3 x = 4\pi r^2dr/ \sqrt{\left(1-2G M(r)/rc^2\right)} $ and not the whole set of Einstein's equations. I identified the Lagrange multiplier $\tilde{\beta}=1/k\tilde{T}$ as the, so-called, Tolman temperature, which corresponds to the surface temperature as measured by an observer at infinity.
Tolman \cite{Tolman:1930} was the first to discover that the local temperature is not constant in equilibrium if general relativity is taken into account. Thermal energy rearranges itself in order to balance its own gravitational attraction \cite{Tolman-Ehrenfest:1930}. This results in a local temperature gradient at equilibrium, and therefore, it is not obvious which is the thermodynamically-conjugate quantity of energy. I have shown, in agreement with Tolman's calculation, not only that the conjugate to energy is the quantity $\tilde{T}$ related to the local temperature by the equation:
\begin{equation}\label{eq:Tolman_T}
	\tilde{T} = T(r) \sqrt{g_{tt}} = const., 
\end{equation}
but in addition to his calculation, I determined $g_{tt}$:
\begin{equation}\label{eq:gtt}
	g_{tt} = e^ {-2 \int_r^\infty dr \left( \frac{G\hat{M}}{\bar{r}^2} + 4\pi G \frac{P}{c^2}\bar{r}\right)
	\left( 1 - \frac{2G\hat{M}}{\bar{r}c^2} \right)^{-1} }
\end{equation}
from the maximum entropy principle \cite{Roupas_RGI_2018}. In the same calculation, it was also proven that $\alpha = \mu(r)/kT(r)=const.$, which leads to Klein's \cite{Klein:1949} relation if combined with (\ref{eq:Tolman_T}).

\begin{figure}[h]
%\begin{center}
\centering
 	\subfigure[The double spiral]{\label{fig:Spiral_double}
 		\includegraphics[scale = 0.4]{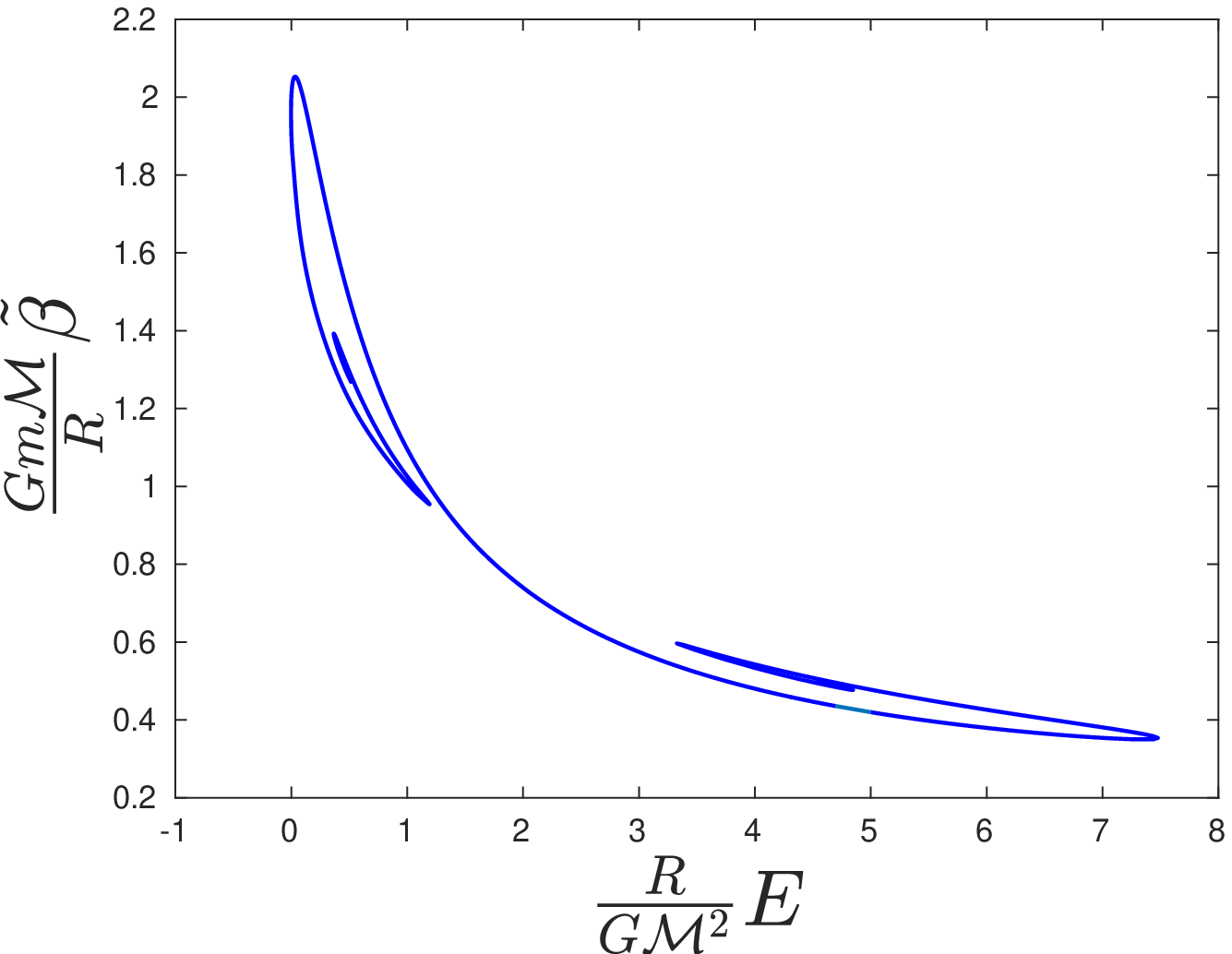} }
	\subfigure[Relativistic gravothermal instability]{ \label{fig:CV_low-spiral}
		\includegraphics[scale = 0.4]{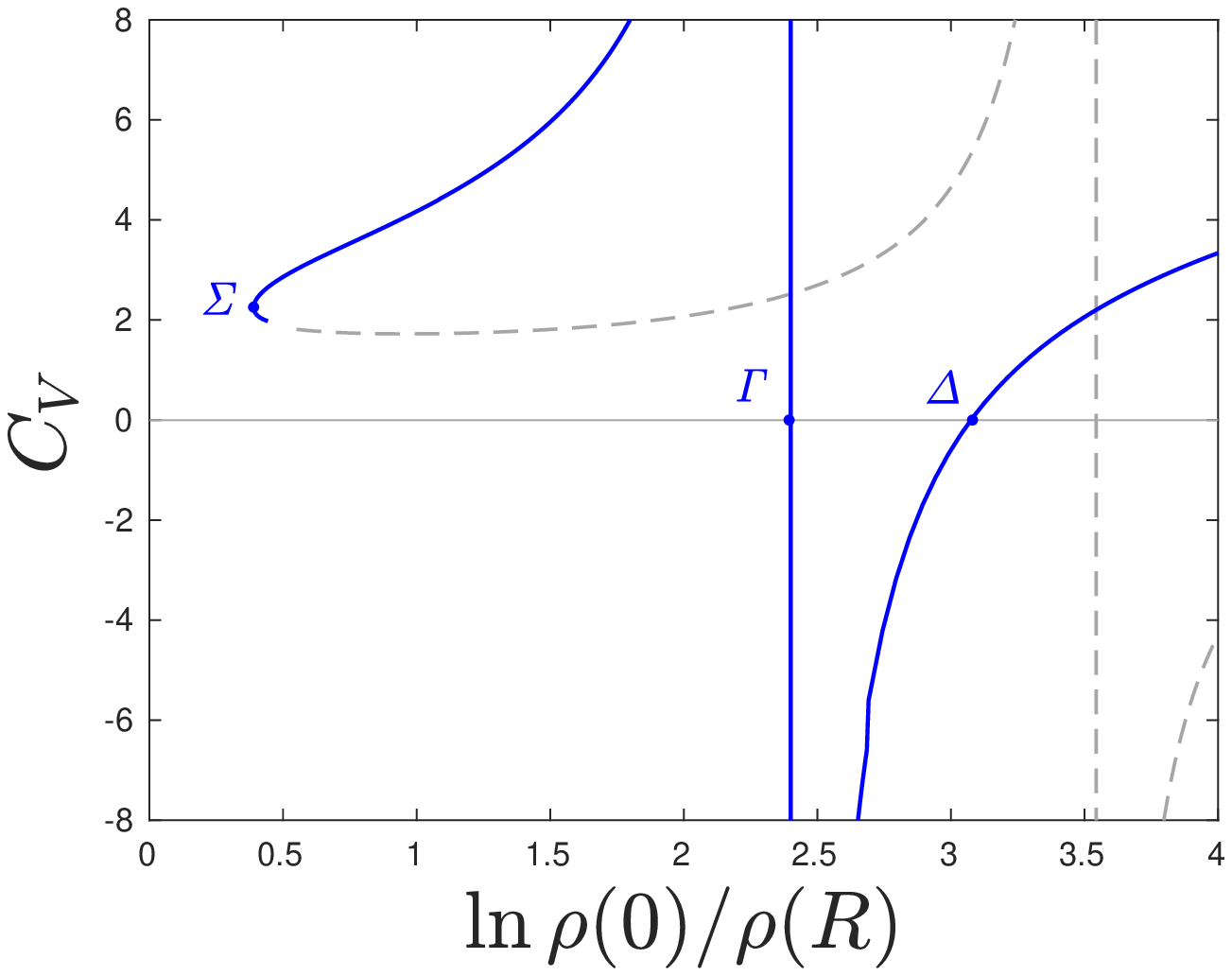} }
	\caption{ {(a):} The caloric curve of the classical ideal gas for rest mass compactness $\xi=0.25$. The~upper spiral designates the low-energy gravothermal instability and the lower spiral the high-energy gravothermal instability. The figure is taken from \cite{Roupas_CQG_RGI_2015}.
	{(b):} The specific heat w.r.t. the density contrast for $\xi=0.1$. The high-energy gravothermal instability sets in at point ${\Delta}$ when the specific heat passes from negative, corresponding to stable equilibria, to positive values. At point ${\Sigma}$, the density contrast attains a minimum. The dashed curve corresponds to the spiral of the low-energy gravothermal instability, that is the upper spiral of the left panel. The figure is taken from \cite{Roupas_RGI_2018}.
	\label{fig:spirals}}
%\end{center} 
\centering
\end{figure}

Now, imagine that we compress the self-gravitating, classical ideal gas sphere. The gas heats up, and the mass density profile gets less steep, tending to a constant density. However, we will reach a point (called ${\Sigma}$ in Figure \ref{fig:spirals}b) beyond which any compression causes a steepening of the mass density profile. Relativity has started to operate. Thermal mass, i.e., the kinetic energy of the random movement of particles, concentrates at the center in order to generate a density gradient to counterbalance {its own gravitational attraction}. The thermal core becomes bound by its own gravity, and eventually, we reach a point (called ${\Delta}$ in Figure \ref{fig:spirals}b) when an instability sets in, the {high-energy gravothermal instability}. The core decouples from the outer regions---the halo---and collapses due to a runaway heat transfer from the core to the halo. 

The caloric curve, i.e., the temperature vs. energy diagram, attains the form of a double spiral, as designated in Figure \ref{fig:spirals}a and discovered for the first time in \cite{Roupas_CQG_RGI_2015}. The upper spiral corresponds to the low-energy gravothermal instability and the lower spiral to the high-energy
gravothermal instability. The intensity of relativistic effects is controlled by the rest mass compactness:
\begin{equation}
\xi = \frac{2GmN}{Rc^2}.
\end{equation}
As $\xi$ is increased, the stability domain is decreased, and the two spirals approach each other. They merge to a single point at $\xi_{max} = 0.35$, above which no equilibria exist.

\begin{figure}[h]
%\begin{center}
\centering
 	\subfigure[Critical radius]{\label{fig:Rcr_micro}
 		\includegraphics[scale = 0.4]{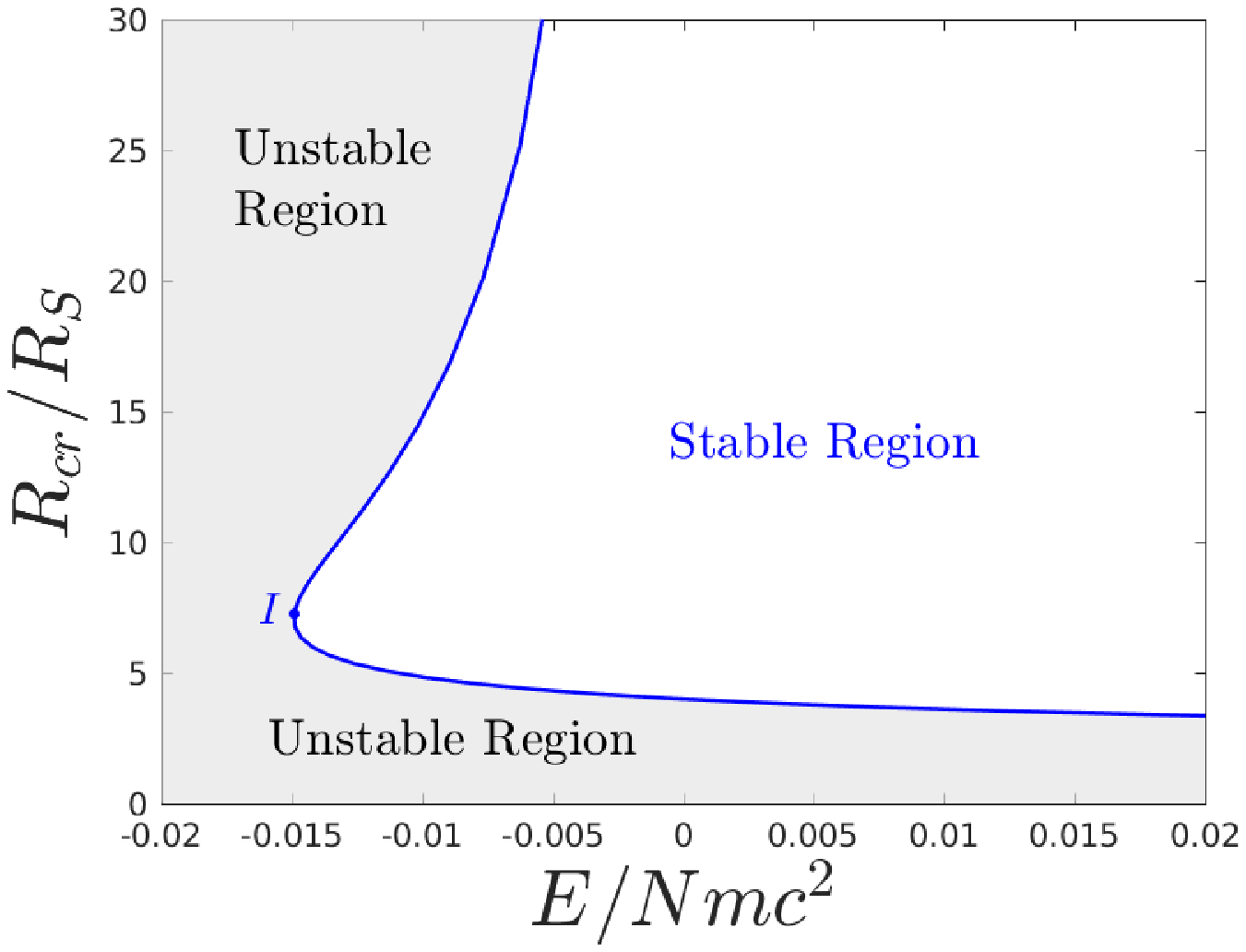} }
 	\subfigure[Critical mass and compactness]{\label{fig:Mcr_micro}
 		\includegraphics[scale = 0.4]{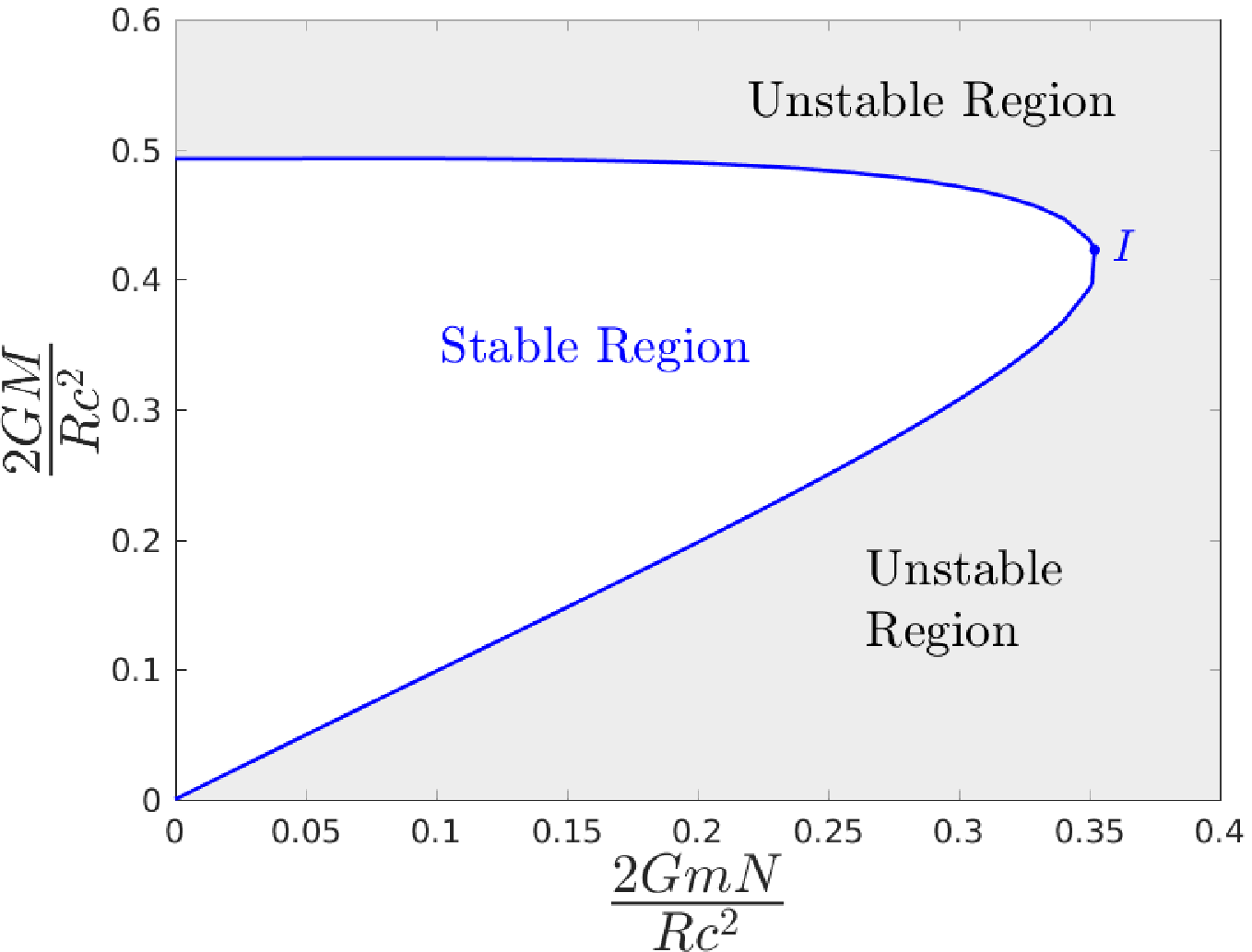} }
	\caption{{(a):} The stability domain of scales for a classical, self-gravitating ideal gas under constant energy conditions. {For a negative fixed value of gravothermal energy higher than} {$-0.015 Nmc^2$}, %is the italic necessary?//ZR: Yes, the math mode is necessary for this expression.
	there exist two marginal radii, in between which stable equilibria do exist. The minimum corresponds to the high-energy gravothermal instability, while the maximum to the low-energy one. The figure is taken from \cite{Roupas_RGI_2018}. {(b):} The stability domain of compactness. For every fixed rest mass compactness, there exist two marginal mass or compactness values, in between which stable equilibria do exist. The maximum corresponds to the high-energy gravothermal instability and the minimum to the low-energy one. No~equilibria exist for rest compactness above $0.35$ and compactness above $0.5$ under any conditions. The~figure is taken from \cite{Roupas_RGI_2018}.
	\label{fig:critical}}
%\end{center} 
\end{figure}

If the gravothermal energy: 
\begin{equation}
	E = Mc^2-Nmc^2
\end{equation}
is negative, but sufficiently high so that stable equilibria do exist, there appear {two critical radii}, which delimit the stable domain shown in Figure \ref{fig:critical}a. The maximum radius corresponds to the (Antonov-type) low-energy gravothermal instability, and the minimum radius signifies the relativistic gravothermal instability. 
The two critical radii merge at the ultra minimum gravothermal energy {$E_\text{min}=-0.015Nmc^2$}, %is the italic necessary for ``Nmc''?//ZR: The math mode is necessary
 below which no equilibria exist. For positive gravothermal energy $E$, there appears only the relativistic gravothermal instability.

In Figure \ref{fig:critical}b is depicted the stable domain w.r.t. the compactness of the system, namely the energy over the radius. In a core-collapse supernova (e.g., \cite{Janka_2012PTEP.2012aA309J,Janka_2012ARNPS..62..407J,Burrows_2013RvMP...85..245B}), the core of a dead star collapses and is heated up during the collapse. It may be halted at densities above the normal nuclear density with a bounce. If the core at the time of the bounce lies in the unstable domain of Figure \ref{fig:critical}b, the high-energy gravothermal instability will set in, and collapse will not be halted \cite{Roupas_RGI_2018}.

\unskip

\section{Mass Limit of Thermal Neutron Cores}

Let us introduce now quantum effects and especially focus on the case of fermions. The equation of state of a relativistic Fermi, ideal gas at any temperature is defined by the set of equations (see, e.g.,~\cite{chandrabook}):
\begin{align}
\label{eq:P_Q} 	
P &= \frac{4\pi g_s m^4 c^5}{3h^3}\int_0^\infty \frac{\sinh^4\theta
d\theta}{e^{b \cosh\theta - \alpha}+1},
\\
\label{eq:rho_Q}
\rho &= \frac{4\pi g_s m^4 c^3}{h^3}\int_0^\infty \frac{\sinh^2\theta
\cosh^2\theta d\theta}{e^{b \cosh\theta - \alpha}+1},
\\
\label{eq:n_Q}
n &= \frac{4\pi g_s m^3 c^3}{h^3}\int_0^\infty \frac{\sinh^2\theta \cosh\theta d\theta}{e^{b \cosh\theta - \alpha}+1}
,
\end{align}
where we denote $b(r)=mc^2/kT$, $\rho=M/V$ the total (relativistic) mass density of the system and $n=N/V$ the rest mass density (consult also \cite{Roupas_2015PRD}). In \cite{Roupas+Chavanis_2018}, we have shown that the high-energy gravothermal instability persists in the case of fermions and under any control parameters' values, revealing its universal character.

Oppenheimer and Volkoff \cite{ov} determined the maximum mass $M_{OV} = 0.7M_\odot$ of an ideal neutron core at zero temperature $T=0$ considering a zero pressure as the boundary condition.
{This mass limit turned out to be very low compared to observations and was later, following the discovery of the first neutron star \cite{Hewish:1968}, regarded as a proof of the fact that nuclear forces have repulsive effects at supra-nuclear densities \cite{Cameron:1959,Zeldovich:1962}. When nuclear forces are taken into account phenomenologically, the limit increases and may exceed two solar masses for cold cores depending on the model (see for example \cite{Haensel:2007,Lattimer:2012} and the references therein), satisfying the latest observational constraint of $2M_\odot$~\mbox{\cite{Demorest:2010,Antoniadis:2013pzd}}.}

In \cite{Roupas_2015PRD}, I generalized to all temperatures the original
calculation of Oppenheimer and Volkoff {for ideal neutron cores. Like them, I neglected nuclear forces. My motivation is partly purely theoretical and partly astrophysical/phenomenological. Regarding the former, consider that if we wished to strip matter from all properties besides these that would allow us to identify it as such, this would result, in the case of fermions, in three basic properties, mass, movement, and quantum degeneracy; that is, gravity, heat, and quantum mechanics. These elementary properties of matter are of particular interest regarding stability, since they antagonize each other over dominating the dynamics of the system. It is not a priori apparent why gravitational instability should exist at all, since the contraction of a system heats it up, adding a stabilizing repulsive force in the form of pressure/thermal energy, while in addition, quantum stabilizing pressure increases. 
{Collapse is generated by the energy loss due to some heat transfer, either external or internal. For example the zero temperature calculations assume implicitly an external heat bath that fixes the temperature at $T=0$. On the other hand, in the case of fixed energy conditions, the system splits inside to a core and a halo generating an internal runaway heat transfer from the former to the latter, which acts as a heat bath (note that in case of relativistic systems the core's gravitation is dominated by the gravitation of thermal energy).}
%either the presence of an external heat bath that will absorb heat (for example the zero temperature calculations assume implicitly a heat bath that fixes the temperature at $T=0$) or, in the case of fixed energy conditions, the splitting of the system to a core and a halo which generates an internal runaway heat transfer (in case of relativistic systems the core's gravitation is dominated by the gravitation of thermal energy).}
%Either the presence of an external heat bath that will absorb heat (for example, the zero temperature calculations assume implicitly a heat bath that fixes the temperature at $T=0$) or, in the case of fixed energy conditions, the gravitational effect of thermal energy causes the collapse (while for weak-field, Newtonian, systems, a halo acts as a heat bath for the contracting core as in gravothermal catastrophe, sometimes called core-collapse). 
Regarding the astrophysical/phenomenological importance of studying the ideal, self-gravitating gas at all temperatures, I stress that the equation of state at supra-nuclear densities, at any temperature, is untraceable by modern experiments and as such only speculative. Therefore, the determination of the behavior of the ideal system at all temperature regimes is crucial. For example, when supplemented by observational data, it may allow us to identify the temperature and mass values at which thermal energy should become dominant over the nuclear forces (whose precise behavior is largely unknown) for the stabilizing of the system. As we will see, interestingly, this occurs at masses $\sim 1.6M_\odot$ and neutron temperatures of some tens of MeV. This also suggests that at these high temperatures, the system may closely resemble an ideal gas. Nevertheless, the~calculation with the equations of state that satisfy the observational constraints at low $T$ would be a natural and very important extension.} 

\begin{figure}[h]
%\begin{center}
\centering
 	\subfigure[Maximum mass and compactness]{\label{fig:M+C_max_thermal_OV}
 		\includegraphics[scale = 0.4]{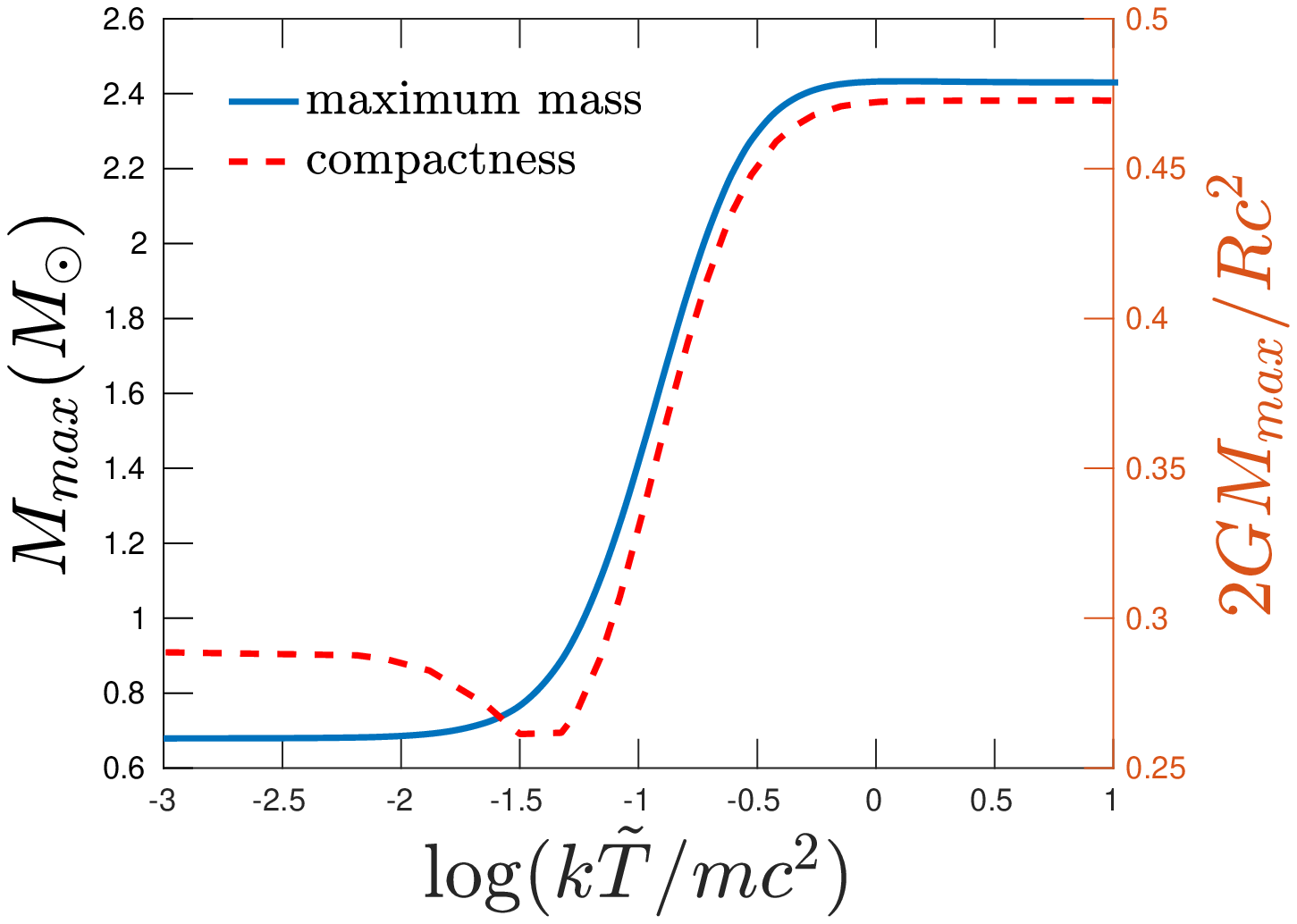} }
 	\subfigure[Maximum mass and corresponding rest mass]{\label{fig:M+Mrest_max_thermal_OV}
 		\includegraphics[scale = 0.4]{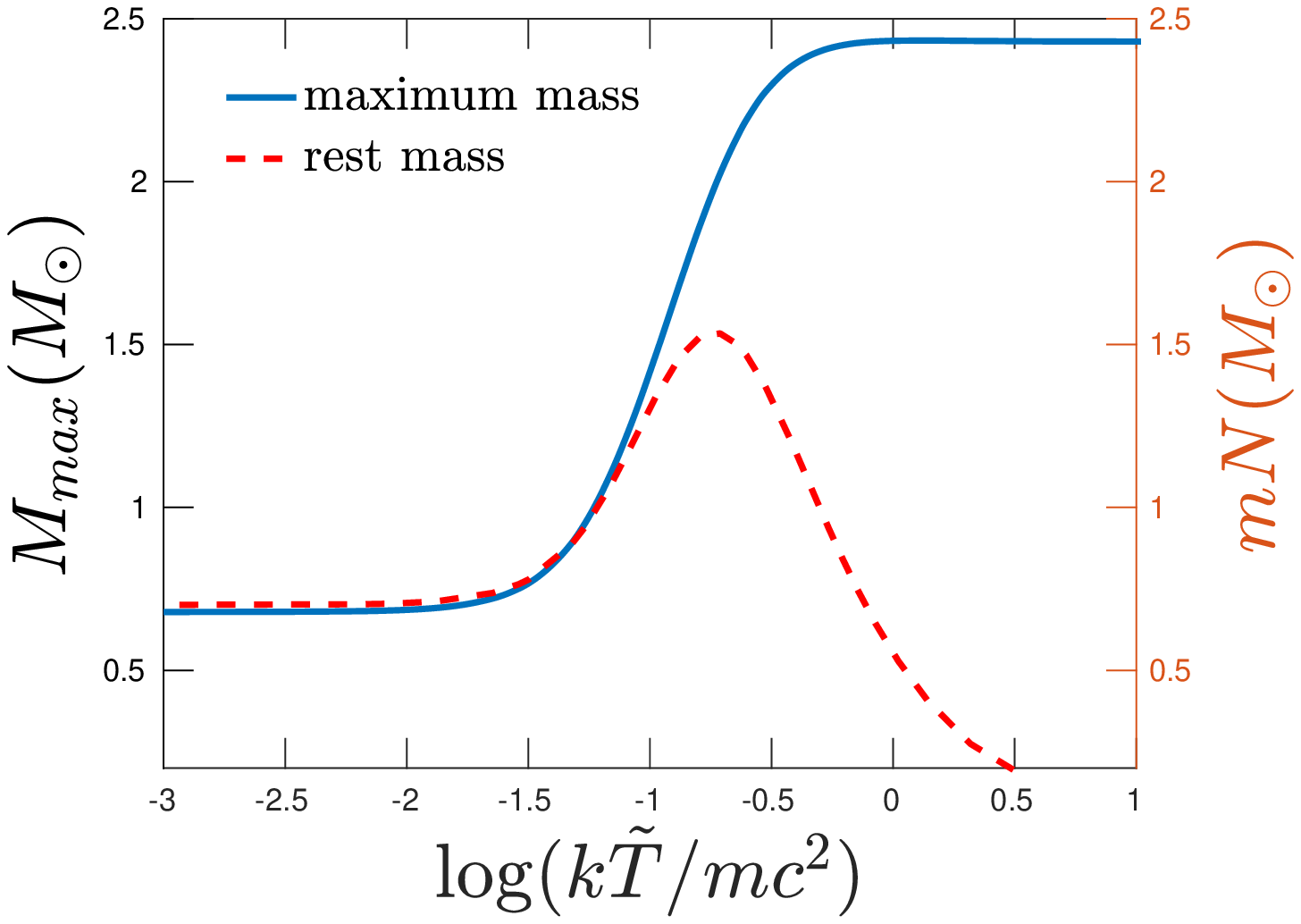} }
	\caption{The maximum mass limit of {ideal} neutron cores w.r.t. temperature. On the left ({\bf a}) is plotted also the corresponding compactness (red dashed line), whose scale is depicted on the right y-axis. In the right panel ({\bf b}) is shown, together with maximum mass, the corresponding rest mass (red dashed line) at the same scale.
	\label{fig:mass_limit}}
%\end{center} 
\end{figure}

{I considered a purely neutron core} and demanded the boundary density of the system to be the maximum density at which heavy nuclei can exist, equivalently the minimum density for which the core is populated dominantly by neutrons. Namely, I assumed $\rho_R = 1.4\times 10^{14}$ gr/cm$^3$, that is half the normal nuclear density. Then, I calculated the mass-radius profiles and determined the maximum mass for several temperature values. The result is plotted in Figure \ref{fig:mass_limit}. At low $T$. the Oppenheimer--Volkoff limit ($0.7M_\odot$) is recovered, while at high $T$, the classical limit ($2.4M_\odot$) \cite{Roupas_CQG_RGI_2015} is recovered.

This calculation fills an apparent gap in the literature. It presents the efficiency of only degeneracy and neutrons' thermal pressures to resist gravity, neglecting other more involved effects. It corresponds to the whole cooling stage of the core starting from the moment of the bounce of a core-collapse supernova incident down to the final cold neutron star. The curve of Figure \ref{fig:mass_limit}a may serve as the idealized framework by which more realistic, involved models (nuclear forces, neutrinos, etc.) of hot cores \cite{Prakash:1997,Lattimer+Prakash_2016PhR...621..127L} can be compared. This allows measuring the efficiency of solely other effects, which will add up to the, now completely determined, primary ones of quantum degeneracy and thermal energy. 

It also reveals that nuclear forces inside the hot core \cite{Prakash:1997,Lattimer+Prakash_2016PhR...621..127L} play a dominant role in stabilizing it up to temperatures of few tens of {MeV}. %is the italic necessary? we advise to remove the italic format for units//ZR: Ok
 I find that {neutrons'} degeneracy and thermal pressures alone can stabilize a thermal core of mass $1.4M_\odot$ only above neutron temperatures $\sim$ 50~{MeV} or more, and therefore nuclear, forces should necessarily intervene to balance the core at these temperatures. Cores of mass {$1.4$--$2.0M_\odot$} %should the minus sign be en dash?
 can be stabilized solely by neutrons' thermal pressure at neutron temperatures of 50--180 {MeV}. Thus, if the core at the moment of the bounce reaches such temperatures, it is stabilized primarily by thermal pressure. However, as we have seen in the previous section, the compactness should be within the stable domain of Figure \ref{fig:critical}b, 
 or else, the core will undergo a high-energy relativistic instability and collapse. Notice that the rest mass presents a maximum at $M_{max}=2M_\odot$ for which $M_\text{rest} = 1.6M_\odot$ (Figure \ref{fig:mass_limit}b), corresponding to a temperature of 180~{MeV} and a radius {of 14.4~km}. It is intriguing that the mass and radius values are close to observations. However, the temperature values are overestimated mainly because neutrino trapping is not taken into account. 

{Figure \ref{fig:mass_limit} reveals the behavior of ideal neutron cores at all temperatures and the intervention between quantum degeneracy pressure, thermal energy, and gravity. The generalization of Figure~\ref{fig:mass_limit} with a more involved equation of state that would incorporate phenomenologically-nuclear forces and satisfy observational constraints at low $T$ would signify an important progress in the field and certainly should be considered in the future.}

\section{Conclusions}

Assume a spherical, classical ideal gas whose constituents mutually interact through gravity at a negative, fixed, total (thermal $+$ gravitational) energy so that it is bound primarily by self-gravity. Nevertheless, some small external boundary pressure is needed to prevent it from slowly evaporating, which acts as an adiabatic wall. Assume further that the radius of the gas sphere is such that the system is thermodynamically stable. For example, it is slightly smaller than $0.335GM^2/|E|$. If the sphere is, adiabatically (with no heat exchange with the environment), expanded above this value, it becomes unstable and collapses because the cooling due to the expansion diminishes the ability of thermal energy to counterbalance the self-gravity of the sphere. However, if the sphere continues to expand, the repelling effect of dark energy will start gradually to become important. Above another threshold radius, the entropy maxima are restored, and stable equilibrium states reappear. This radius, from the point of view of an expanding Universe, signifies the turnaround radius, i.e., the radius below which it is possible for structures to form due to gravitational collapse. This radius is bigger for a quintessential dark energy and smaller for phantom dark energy w.r.t. the cosmological constant. Thus, quintessence favors the formation of structure, since it maximizes the instability domain.

Consider now the opposite situation, where the sphere starts to contract adiabatically. Initially, the~mass density contrast, namely the central to edge mass density ratio, is decreasing. The mass of the gas becomes progressively more homogeneously distributed due to the increase of temperature. Thermal energy's ability to stabilize gravity is enhancing. However, there exists a radius below which the mass density contrast starts to increase while the gas is contracted, rendering the gas less homogeneously distributed!%is the exclamation necessary?
 The reason is that relativistic effects have started to become evident. The~mass of thermal motion accumulates in the center. Now, there appears a minimum radius threshold below which the gas becomes unstable due to the weight of heat. 

This effect, the {high-energy relativistic gravothermal instability}, persists even if the quantum degeneracy pressure of fermionic particles is taken into account. In particular, for neutrons in Figure \ref{fig:critical}b, the stable domain is depicted w.r.t. the compactness of the system, namely the energy over the radius. In~a core-collapse supernova, the core of a dead star collapses and is heated up during the collapse. It may be halted at densities above the normal nuclear density with a bounce. If the core at the time of the bounce lies in the unstable domain of Figure \ref{fig:critical}b, the high-energy gravothermal instability will set in, and the collapse will not be halted. 
I find that neutrons' degeneracy and thermal pressures alone can stabilize a neutron core of mass 1.4--2.0$M_\odot$ at neutron temperatures of {50--180 ~MeV}, corresponding to an ultra-hot proto-neutron star. However, nuclear forces inside the hot core are expected to play a dominant role in stabilizing it up to a temperature of a few tens of MeV.

%\vspace{6pt} 
%\authorcontributions{\hl{For research articles with several authors}, 
%a short paragraph specifying their individual contributions must be provided. The following statements should be used ``conceptualization, X.X. and Y.Y.; methodology, X.X.; software, X.X.; validation, X.X., Y.Y. and Z.Z.; formal analysis, X.X.; investigation, X.X.; resources, X.X.; data curation, X.X.; writing--original draft preparation, X.X.; writing--review and editing, X.X.; visualization, X.X.; supervision, X.X.; project administration, X.X.; funding acquisition, Y.Y.'', please turn to the \href{http://img.mdpi.org/data/contributor-role-instruction.pdf}{CRediT taxonomy} for the term explanation. Authorship must be limited to those who have contributed substantially to the work reported.}

%%%%%%%%%%%%%%%%%%%%%%%%%%%%%%%%%%%%%%%%%%

%\externalbibliography{yes}
\bibliography{ROUPAS_Zacharias_ICFNP2018}

\begin{thebibliography}{10}

\bibitem{antonov}
V.~A. {Antonov}.
\newblock {\em Solution of the problem of stability of stellar system with
  Emden's density law and the spherical distribution of velocities\/}.
\newblock Vestnik Leningradskogo Universiteta, Leningrad: University, 1962.

\bibitem{lbw}
D.~{Lynden-Bell} and R.~{Wood}.
\newblock {The gravo-thermal catastrophe in isothermal spheres and the onset of
  red-giant structure for stellar systems}.
\newblock {\em \mnras\/}, 138:495, 1968.

\bibitem{2013A&A...552A..37S}
M.~C. {Sormani} and G.~{Bertin}.
\newblock {Gravothermal catastrophe: The dynamical stability of a fluid model}.
\newblock {\em \aap\/}, 552:A37, April 2013.

\bibitem{Jeans_1902RSPTA.199....1J}
J.~H. {Jeans}.
\newblock {The Stability of a Spherical Nebula}.
\newblock {\em Philosophical Transactions of the Royal Society of London Series
  A\/}, 199:1--53, 1902.

\bibitem{Tolman:1930}
Richard~C. Tolman.
\newblock On the weight of heat and thermal equilibrium in general relativity.
\newblock {\em Phys. Rev.\/}, 35:904, 1930.

\bibitem{Roupas_RGI_2018}
Z.~{Roupas}.
\newblock {Relativistic Gravothermal Instability: the Weight of Heat}.
\newblock {\em ArXiv e-prints\/}, September 2018.

\bibitem{Roupas_CQG_RGI_2015}
Z.~{Roupas}.
\newblock {Relativistic gravothermal instabilities}.
\newblock {\em Class. Quant. Grav.\/}, 32(13):135023, 2015.

\bibitem{Axenides_etal_2012PRD}
M.~{Axenides}, G.~{Georgiou}, and Z.~{Roupas}.
\newblock {Gravothermal catastrophe with a cosmological constant}.
\newblock {\em \prd\/}, 86(10):104005, November 2012.

\bibitem{Axenides_2013NuPhB.871...21A}
M.~{Axenides}, G.~{Georgiou}, and Z.~{Roupas}.
\newblock {Gravitational instabilities of isothermal spheres in the presence of
  a cosmological constant}.
\newblock {\em Nuclear Physics B\/}, 871:21--51, June 2013.

\bibitem{Roupas_etal_2014PRD}
Z.~{Roupas}, et~al.
\newblock {Galaxy clusters and structure formation in quintessence versus
  phantom dark energy universe}.
\newblock {\em \prd\/}, 89(8):083002, April 2014.

\bibitem{Stuchlik_1999PhRvD..60d4006S}
Z.~{Stuchl{\'{\i}}k} and S.~{Hled{\'{\i}}k}.
\newblock {Some properties of the Schwarzschild-de Sitter and
  Schwarzschild-anti-de Sitter spacetimes}.
\newblock {\em \prd\/}, 60(4):044006, August 1999.

\bibitem{Busha_2003ApJ...596..713B}
M.~T. {Busha}, et~al.
\newblock {Future Evolution of Cosmic Structure in an Accelerating Universe}.
\newblock {\em \apj\/}, 596:713--724, October 2003.

\bibitem{Mizony_2005A&A...434...45M}
M.~{Mizony} and M.~{Lachi{\`e}ze-Rey}.
\newblock {Cosmological effects in the local static frame}.
\newblock {\em \aap\/}, 434:45--52, April 2005.

\bibitem{Pavlidou_2014JCAP...09..020P}
V.~{Pavlidou} and T.~N. {Tomaras}.
\newblock {Where the world stands still: turnaround as a strong test of
  {$\Lambda$}CDM cosmology}.
\newblock {\em \jcap\/}, 9:020, September 2014.

\bibitem{Faraoni_2015JCAP...10..013F}
V.~{Faraoni}, M.~{Lapierre-L{\'e}onard}, and A.~{Prain}.
\newblock {Turnaround radius in an accelerated universe with quasi-local mass}.
\newblock {\em \jcap\/}, 10:013, October 2015.

\bibitem{Capozziello_2018arXiv180501233C}
S.~{Capozziello}, K.~F. {Dialektopoulos}, and O.~{Luongo}.
\newblock {Maximum turnaround radius in $f(R)$ gravity}.
\newblock {\em ArXiv e-prints\/}, May 2018.

\bibitem{Roupas_2013CQG}
Z.~{Roupas}.
\newblock {Thermodynamical instabilities of perfect fluid spheres in General
  Relativity}.
\newblock {\em Classical and Quantum Gravity\/}, 30(11):115018, June 2013.

\bibitem{Roupas_2015CQG_32k9501R}
Z.~{Roupas}.
\newblock {Corrigendum: Thermodynamical instabilities of perfect fluid spheres
  in General Relativity}.
\newblock {\em Classical and Quantum Gravity\/}, 32(11):119501, June 2015.

\bibitem{Tolman-Ehrenfest:1930}
Richard~C. Tolman and Paul Ehrenfest.
\newblock Temperature equilibrium in a static gravitational field.
\newblock {\em Phys. Rev.\/}, 36:1791--1798, 1930.

\bibitem{Klein:1949}
O.~{Klein}.
\newblock {On the Thermodynamical Equilibrium of Fluids in Gravitational
  Fields}.
\newblock {\em Reviews of Modern Physics\/}, 21:531--533, July 1949.

\bibitem{Janka_2012PTEP.2012aA309J}
H.-T. {Janka}, et~al.
\newblock {Core-collapse supernovae: Reflections and directions}.
\newblock {\em Progress of Theoretical and Experimental Physics\/},
  2012(1):01A309, December 2012.

\bibitem{Janka_2012ARNPS..62..407J}
H.-T. {Janka}.
\newblock {Explosion Mechanisms of Core-Collapse Supernovae}.
\newblock {\em Annual Review of Nuclear and Particle Science\/}, 62:407--451,
  November 2012.

\bibitem{Burrows_2013RvMP...85..245B}
A.~{Burrows}.
\newblock {Colloquium: Perspectives on core-collapse supernova theory}.
\newblock {\em Reviews of Modern Physics\/}, 85:245--261, January 2013.

\bibitem{chandrabook}
S.~Chandrasekhar.
\newblock {\em An introduction to the study of stellar structure\/}.
\newblock Chicago, iLLINOIS, 1938.

\bibitem{Roupas_2015PRD}
Z.~{Roupas}.
\newblock {Thermal mass limit of neutron cores}.
\newblock {\em \prd\/}, 91(2):023001, January 2015.

\bibitem{Roupas+Chavanis_2018}
Zacharias Roupas and Pierre-Henri Chavanis.
\newblock {Relativistic Gravitational Phase Transitions and Instabilities of
  the Fermi Gas}.
\newblock {\em ArXiv e-prints\/}, 2018.

\bibitem{ov}
J.~R. {Oppenheimer} and G.~M. {Volkoff}.
\newblock {On Massive Neutron Cores}.
\newblock {\em Physical Review\/}, 55:374--381, February 1939.

\bibitem{Hewish:1968}
A.~Hewish, et~al.
\newblock Observation of a rapidly pulsating radio source.
\newblock {\em Nature\/}, 709, 1968.

\bibitem{Cameron:1959}
A.~Cameron.
\newblock Pycnonuclear reactions and nova explosions.
\newblock {\em Astrophys. J.\/}, 130:916, 1959.

\bibitem{Zeldovich:1962}
Y.~Zeldovich.
\newblock {\em Sov. Phys.-JETP\/}, 14:1143, 1962.

\bibitem{Haensel:2007}
P.~Haensel, A.~Potekhin, and D.~Yakovlev.
\newblock {\em Neutron Stars I\/}.
\newblock Springer, 2007.

\bibitem{Lattimer:2012}
J.M. Lattimer.
\newblock The nuclear equation of state and neutron star masses.
\newblock {\em Annual Review of Nuclear and Particle Science\/}, 62:485, 2012.

\bibitem{Demorest:2010}
Paul Demorest, et~al.
\newblock {Shapiro delay measurement of a two solar mass neutron star}.
\newblock {\em Nature\/}, 467:1081--1083, 2010.

\bibitem{Antoniadis:2013pzd}
John Antoniadis, et~al.
\newblock {A Massive Pulsar in a Compact Relativistic Binary}.
\newblock {\em Science\/}, 340:6131, 2013.

\bibitem{Prakash:1997}
Madappa Prakash, et~al.
\newblock {Composition and structure of protoneutron stars}.
\newblock {\em Physics Reports\/}, 280:1--77, 1997.

\bibitem{Lattimer+Prakash_2016PhR...621..127L}
J.~M. {Lattimer} and M.~{Prakash}.
\newblock {The equation of state of hot, dense matter and neutron stars}.
\newblock {\em \physrep\/}, 621:127--164, March 2016.

\end{thebibliography}
\bibliographystyle{myunsrt}

\end{document}